\documentclass[a4paper,flushrt,twocolumn]{aastex62}
\usepackage{amsmath, amsfonts, amssymb, graphicx, rotating, natbib}
\usepackage{epstopdf}
\DeclareGraphicsExtensions{.png,.pdf}
\bibpunct{(}{)}{;}{a}{}{,}

\newcommand{\kms}{km~s$^{-1}$}

\shorttitle{ISN Gas And PUIs: Ionization}
\shortauthors{Sok{\'o\l} et al.}
\begin{document}
%
\title{\textbf{Interstellar Neutral Gas Species and Their Pickup Ions inside the Heliospheric Termination Shock. Ionization Rates for H, O, Ne, and He}}
\correspondingauthor{Justyna M. Sok{\'o\l}}
\email{jsokol@cbk.waw.pl}
\author{Justyna M. Sok{\'o\l}}
\affiliation{Space Research Centre, Polish Academy of Sciences, (CBK PAN), Warsaw, Poland}
\author{Maciej Bzowski}
\affiliation{Space Research Centre, Polish Academy of Sciences, (CBK PAN), Warsaw, Poland}
\author{Munetoshi Tokumaru}
\affiliation{Institute for Space-Earth Environmental Research, Nagoya University, Nagoya, Japan}
%
%
\begin{abstract}
Solar ionizing factors are responsible for modulation of interstellar neutral gas and its derivative populations inside the heliosphere. We provide an overview of the current state of knowledge about them for heliospheric particles inside the termination shock. We discuss charge exchange with solar wind particles, photoionization, and electron impact ionization for hydrogen, oxygen, neon, and helium from 1985 to 2018 both in the ecliptic plane and in the polar regions. We discuss ionization rates as a function of time, distance to the Sun, and latitude. We compare the total ionization rates among the species within a consistent and homogeneous system of calculation of the ionization rates. The highest total ionization rates at 1~au in the ecliptic plane are for hydrogen and oxygen, and the lowest are for helium. In the polar regions, the strongest ionization losses are for oxygen, regardless of the solar activity. Photoionization is the dominant ionization reaction for helium and neon, and a reaction of high significance for oxygen. Charge exchange with solar wind particles is the dominant ionization reaction for hydrogen and the second important ionization reaction for oxygen. Electron impact ionization is an important ionization reaction for Ne and He, with the contribution to the total ionization rates stronger within 1~au and smaller outside. The total ionization rates for He and Ne vary in time with the solar activity, whereas the total ionization rates for H and O follow the cyclic solar wind variations out of the ecliptic plane and aperiodic variations in the ecliptic plane.  
\end{abstract}


\section{Introduction \label{sec:intro}}

The heliospheric particles, like pickup ions (PUIs) and energetic neutral atoms (ENAs), and heliospheric phenomena, like the resonant backscatter glow in the solar hydrogen (121.6~nm) and helium (58.4~nm) lines, result from interaction of the interstellar medium with the solar output inside the heliosphere. The interstellar neutral (ISN) particles enter the heliosphere unaffected by the magnetic field, and during their travel inside the heliosphere they are exposed to the solar ionizing factors, like collisions with solar wind particles (protons, alphas, and electrons) and solar extreme ultraviolet (EUV) radiation. In consequence of ionization of ISN particles, new populations of particles are created. In general, ionization of ISN gas creates PUIs, charge exchange reactions of PUIs and neutral atoms create various populations of ENAs. ISN gas, PUIs, and ENAs can reach close distances to the Sun and be sampled by heliospheric space probes like, e.g., \textit{IBEX} \citep{mccomas_etal:09a}, \textit{Ulysses} \citep{wenzel_etal:89a,witte_etal:92a}, \textit{New Horizons} \citep{mccomas_etal:08b}, \textit{Cassini} \citep{young_etal:04a}, to list only a few. All of these heliospheric particles interact with the solar output and undergo losses due to ionization processes. A careful accounting for the solar ionization factors is needed to deconvolve the solar modification of the particle fluxes on their way from the source region to the instruments from the modulation in the source region to correctly interpret the measurements \citep{bzowski:08a, mccomas_etal:12c}. Thus, the knowledge of the solar ionization factors and their modulation in time and space is fundamental in the study of the ISN gas distribution, PUI production rate, and the physics of ENAs in the heliosphere.

This paper belongs to the series \emph{Interstellar Neutral Gas Species And Their Pickup Ions Inside The Heliospheric Termination Shock}, in which topics related to the study and modeling of ISN gas and PUIs inside the heliosphere are discussed. Here we summarize the current state of knowledge about the solar ionizing factors inside the heliosphere. We discuss the most relevant ionization processes for heliospheric particles based on a homogeneous system of calculation of ionization losses relevant for ISN gas, PUIs, and ENAs inside the heliosphere developed in our previous studies and available in the literature \citep{rucinski_etal:98, bzowski:08a, bzowski_etal:13a, bzowski_etal:13b, sokol_etal:13a, sokol_etal:15d, sokol_etal:16a}. The system is built on the currently best available solar observations. 

We overview the ionization rates for the four most abundant ISN species inside the heliosphere: hydrogen, helium, oxygen, and neon as a function of time, distance to the Sun, and latitude. We study the solar ionizing factors backward in time as long as observational data are available. We focus on Carrington rotation period\footnote{Carrington rotation is the mean rotation period of the Sun observed from the Earth equal to 27.2753 days.} resolution in time. We study relations of the ionization rates among these species, we discuss similarities and differences and the resulting consequences. 

The paper is organized as follows: a brief description of ionization reactions relevant for the study of heliospheric particles together with references to the currently used models is presented in Section~\ref{sec:ionization}. Relations between total ionization rates for various species are discussed in Section~\ref{sec:totIon}. We summarize the study in Section~\ref{sec:Summary}.

\section{Ionization processes \label{sec:ionization}}
The interaction of heliospheric particles with the solar medium is via charge exchange with solar wind protons and alpha particles ($\beta_{\mathrm{cx}}$, Section~\ref{sec:cx}), photoionization by the solar EUV radiation ($\beta_{\mathrm{ph}}$, Section~\ref{sec:ph}), and the impact ionization by solar wind electrons ($\beta_{\mathrm{el}}$, Section~\ref{sec:ele}). The ionization processes for the ISN gas were discussed in detail by \citet{rucinski_etal:96a, rucinski_etal:98, fahr_etal:07a}, a comprehensive review can be also found in \citet{bzowski_etal:13a} for H and \citet{bzowski_etal:13b} for He, Ne, and O. An overview of the ionization processes relevant for heliospheric particles, however, in the heliosheath, can be found in \citet{scherer_etal:14a}. 

\subsection{Charge exchange \label{sec:cx}}
Charge exchange is a reaction in which a charge (electron) is captured by an ion (in our case proton or alpha particle) from a colliding atom:
\begin{displaymath}
atom_{\mathrm{ISN}} + ion_{\mathrm{sw}} \to PUI + ENA.
\end{displaymath}
The intensity of charge exchange ionization for ISN atoms is a function of concentration of the impacting component, relative velocity of the reacting particles, and the cross-section for the reaction, as given by Equation~\ref{eq:cx}:
\begin{equation}
\beta_{\mathrm{cx}}\left(t,r,\phi\right) = n_{\mathrm{sw}}\left(t,r,\phi\right)v_{\mathrm{rel}}\left(t,r,\phi\right)\sigma_{\mathrm{cx}}\left( v_{\mathrm{rel}} \left( t,r,\phi \right) \right),
\label{eq:cx}
\end{equation}
with 
\begin{displaymath}
v_{\mathrm{rel}}\left ( t,r,\phi \right)= | \vec{v}_\mathrm{atom}\left ( t,r,\phi \right) - \vec{v}_{\mathrm{sw}}\left ( t,r,\phi \right) |,
\end{displaymath}
where $\vec{v}_\mathrm{atom}$ is the velocity of an ISN atom, $\vec{v}_{\mathrm{sw}}$ is the solar wind velocity, assumed to flow radially, $n_{\mathrm{sw}}$ is the concentration of solar wind particles (protons or alpha particles, depending on the reaction), and $\sigma_{\mathrm{sw}}$ is the cross-section for charge exchange. We assumed that the solar wind varies in time ($t$), distance to the Sun ($r$), and heliographic latitude ($\phi$). We adopt a decrease with the square of distance to the Sun for the solar wind density and a lack of dependence on heliocentric distance for the solar wind velocity:
\begin{equation}
n_{\mathrm{sw}}\left( r \right)=n_{\mathrm{sw}}\left(r_0\right)\left(\frac{r_0}{r}\right)^2, \quad
v_{\mathrm{sw}}\left( r \right)=v_{\mathrm{sw}}\left(r_0\right),
\label{eq:swR}
\end{equation}
where $r_0 = 1$~au.

We use the charge exchange cross-sections after \citet{lindsay_stebbings:05a} for H and O, \citet{nakai_etal:87a} for Ne, and \citet{barnett_etal:90} for He. In the case of helium atoms, the charge exchange ionization reaction, in addition to charge exchange with protons, may be due to interaction with the alpha particles ($He + \alpha \to He^+ + He^+$, $He + \alpha \to He^{++} + He$), which we account for in our calculations. The charge exchange rate for He presented further is a sum of the rates with protons and alpha particles.

To calculate the charge exchange rates, we use the model of the solar wind proton speed and density evolution in time and latitude developed by \citet{sokol_etal:13a} with a modification for the retrieval of the solar wind proton density variations with latitude described by \citet{sokol_etal:15d}. The model is based on in-ecliptic multispacecraft solar wind data from the OMNI collection \citep{king_papitashvili:05} and indirect solar wind speed observations via the interplanetary scintillations (IPS, \citep{tokumaru_etal:15a}). More about the solar wind data used can be found in Section~\ref{sec:sw}. For the calculation of charge exchange with alpha particles we adopted a constant fraction of alpha particles in the solar wind equal to $4\%$ of the proton number density. Although, the alpha-to-proton fraction varies significantly in the slow solar wind (from $\sim2\%$ to $\sim7\%$), its variation in the fast solar wind, and thus with latitude, is moderate with solar cycle with an average value of about $4\%$ \citep{kasper_etal:12a}.

\subsection{Photoionization \label{sec:ph}}
Photoionization is an ionization via knock out of an electron from an atom by photons:
\begin{displaymath}
X + h\nu \to X^+ + e.
\end{displaymath}
A particle can be ionized by photons with the energies higher than the ionization threshold energy for a given species: 13.60~eV for H, 24.60~eV for He, 13.62~eV for O, and 21.57~eV for Ne. These energies require photons of wavelengths equal to or smaller than 91.18~nm, 50.43~nm, 91.05~nm, and 57.50~nm, respectively. This makes the EUV spectral range of the solar electromagnetic radiation the main source of photoionization for heliospheric particles.

Photoionization rate is calculated by integration of the solar EUV spectral flux ($F_{\mathrm{EUV}}$) for a given wavelength $\lambda$ multiplied by the cross-section for photoionization by photon of this wavelength ($\sigma_{\mathrm{ph}}$), as given in Equation~\ref{eq:ph}:
\begin{equation}
\beta_{\mathrm{ph}}\left(t\right) = \int\limits_{0}^{\lambda_0} F_{\mathrm{EUV}} \left(\lambda, t\right) \sigma_{\mathrm{ph}}\left( \lambda \right) \mathrm{d}\lambda,
\label{eq:ph}
\end{equation}
where $\lambda_0$ is wavelength for the ionization threshold for a given atom. The cross-sections for photoionization are adopted from \citet{verner_etal:96}. 

The EUV flux is modulated by the distribution of active sources on the solar surface. The latitudinal anisotropy of the EUV flux was studied by \citet{cook_etal:80a, cook_etal:81a} and \citet{pryor_etal:92} and, recently, extensively by \citet{auchere_etal:05a,auchere_etal:05c}, who reported that the pole-to-equator ratio for the He~II~30.4~nm flux varies in a range from 0.6 during solar maximum up to 0.9 during solar minimum. Additionally, \citet{auchere:05a} showed that the latitudinal anisotropy of the chromospheric flux in Lyman-$\alpha$ line varies with distance to the Sun and is found to be 15$\%$ at 1~au during solar minimum. Such variation was also estimated for solar maximum by \citet{pryor_etal:92}. But, on the other hand, \citet{kiselman_etal:11a} reported a weak variation with latitude for various solar lines. Since calculation of the photoionization rates requires integration over broad spectral ranges, a careful study of the latitudinal variation of the solar lines is needed, which is outside the scope of this paper. Thus, for the purpose of our study we follow the assumption made by \citet{bzowski_etal:13a} of a moderate variation in heliographic latitude constant in solar cycle, as given in Equation~\ref{eq:phLat}:
\begin{equation}
 \beta_{\mathrm{ph}}\left(\phi\right)=\beta_{\mathrm{ph}}\left(0\degr\right)\left( 0.85 \sin^2\left( \phi \right) + \cos^2 \left( \phi \right) \right).
\label{eq:phLat}
 \end{equation}
This approximation does not reflect solar cycle variations reported by \citet{auchere_etal:05a}. This needs a separate study, because the data about the pole-to-equator variations are not available for the whole time period we study. A consequence of the assumption of constant and mild latitudinal variation of the photoionization rates may overestimate the photoionization rates for polar regions during solar maximum, especially for distance to the Sun smaller than 1~au, as pointed out by \citet{auchere:05a}.

The heliospheric environment is optically thin for the ionizing photons, thus for the distance dependence of photoionization, we adopt the following relation:
\begin{equation}
\beta_{\mathrm{ph}}\left(r\right)=\beta_0\left( \frac{r_0}{r} \right)^2.
\label{eq:phR}
\end{equation}

To calculate photoionization rates, we used the methodology described by \citet{bzowski_etal:13a, bzowski_etal:13b} to create a composite series of photoionization rates. We used the available data of the solar EUV spectral irradiance from TIMED/SEE measurements \citep[Level~3, Version~11;][]{woods_etal:05a} and a series of solar EUV proxies \citep[see also][]{bochsler_etal:14a,bochsler_etal:14ae}. As the TIMED time series is spectral irradiance given in the units of W\,m$^{-2}$\,nm$^{-1}$, to use it in Equation~\ref{eq:ph} a prior scaling to energy per photon is required. As solar EUV proxies, we used the flux measured by SOHO/CELIAS/SEM \citep{judge_etal:98, wieman_etal:14a}, F10.7 radio flux \citep{tapping:87, tapping:13a}, the magnesium index \citep[Mg$_{c/w}$,][]{snow_etal:14a}, and the composite Lyman-$\alpha$ flux released by LASP \citep{woods_etal:00}. Details of the construction of the photoionization rates can be found in \citet{sokol_bzowski:14a} for He, \citet{bzowski_etal:13b} for Ne and O, and \citet{bzowski_etal:13a} for H.

\subsection{Electron impact ionization \label{sec:ele}}
\begin{figure}
\includegraphics[width=0.45\textwidth]{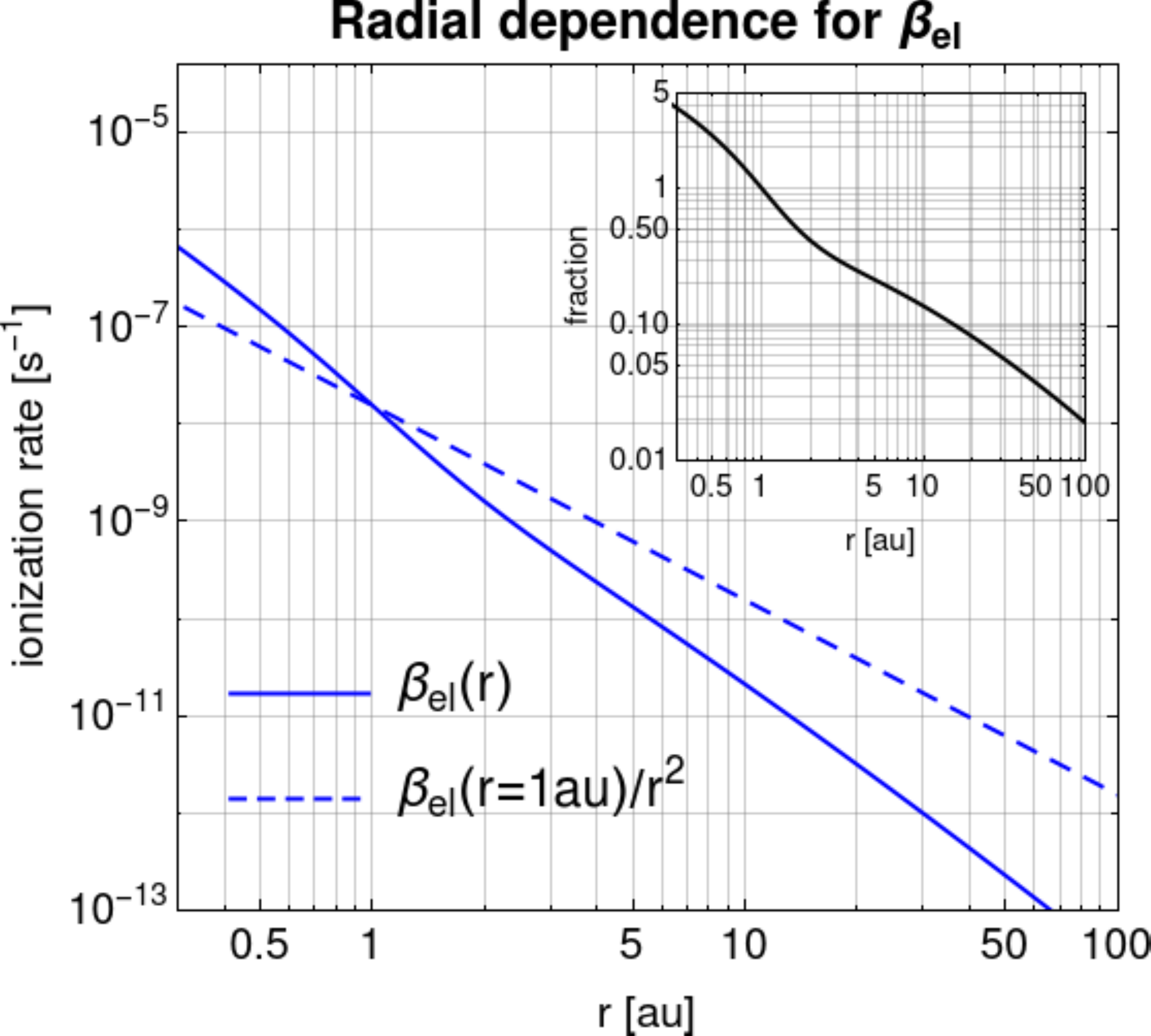}
\centering
\caption{Variation of the electron impact ionization rate with distance to the Sun as an example for He for the slow solar wind for a moderate phase of solar activity as in 1999. The solid line, $\beta_{\mathrm{el}}(r)$, presents the radial dependence obtained using the methodology discussed in Section~\ref{sec:ele} and based on the solar wind electron temperature profiles measured by \textit{Helios} inside 1~au and \textit{Ulysses} inside 5~au. The dotted line illustrates the $r^{-2}$ scaling of the electron impact ionization rate measured at 1~au $\beta_{\mathrm{el}}(r=1~\mathrm{au})$. Inset in the upper right corner shows the departure of electron impact ionization from $r^{-2}$ scaling by illustrating the fraction $\beta_{\mathrm{el}}(r)/\left( \beta_{\mathrm{el}}(r=1)r^{-2} \right)$. \label{fig:eleR}}
\end{figure}

Electron impact ionization is a reaction in which an electron knocks out another electron from an atom:
\begin{displaymath}
X + e \to X^+ + 2e.
\end{displaymath}

The calculation of electron impact ionization rate includes the distribution function of the solar wind electrons  ($f_{\mathrm{el}}\left( E,t,r,\phi \right)$, where $E$ is the collision energy) and the cross-section for electron impact ionization ($\sigma_{\mathrm{el}}$) as given in Equation~\ref{eq:el} after \citet{rucinski_fahr:89}:
\begin{equation}
\beta_{\mathrm{el}}\left(t,r,\phi \right) = \frac{8\pi}{m_{\mathrm{e}}^2}\int\limits_{E_i}^{\infty}f_{\mathrm{el}}\left( E,t,r,\phi \right)\sigma_{\mathrm{el}}\left(E\right)E \mathrm{d}E,
\label{eq:el}
\end{equation}
where $m_{\mathrm{e}}$ is mass of an electron. The cross-sections are adopted after \citet{lotz:67}. In our study we calculate the electron impact ionization following the methodology proposed by \citet{rucinski_fahr:89,rucinski_fahr:91} and developed by \citet{bzowski:08a, bzowski_etal:13b} based on \textit{Helios} measurements inside 1~au \citep{marsch_etal:89} and \textit{Ulysses} measurements inside 5~au \citep{scime_etal:94}. An extrapolation was applied inward and outward the heliocentric distance range investigated by Helios and \textit{Ulysses}. The distribution function is approximated by two Maxwell-Boltzmann functions representing the cool core and the hot halo populations with increasing abundance of the latter one with distance from the Sun. The temperatures and densities of the core and halo populations vary with distance from the Sun differently than $r^{-2}$, as measured by \textit{Helios} and \textit{Ulysses} \citep{scime_etal:94, issautier_etal:98, maksimovic_etal:00a}. Thus, unlike the charge exchange ionization and photoionization, the electron impact ionization decreases with the the distance to the Sun $r$ faster than $r^{-2}$:
\begin{displaymath}
\beta_{\mathrm{el}}\left(r\right)\neq\beta_0\left(\frac{r_0}{r} \right)^2.
\end{displaymath}
This is because the solar wind electrons are not isothermal with heliocentric distance, as presented in Equations~3.31 and 3.32 in \citet{bzowski_etal:13a} based on findings for core and halo electron populations for the slow and fast solar wind flows. Figure~\ref{fig:eleR} compares the distance dependence of electron impact ionization decrease with the $r^{-2}$ decrease in the case of helium.

Details of the electron impact ionization for ISN~H are given by \citet{bzowski:08a} and for ISN~He, Ne, and O by \citet{bzowski_etal:13b}, and extensive discussion is also presented by \citet{bzowski_etal:13a}. In the calculation on the electron impact ionization we use the solar wind density variations in time and latitude based on the model by \citet{sokol_etal:13a} with further modifications discussed by \citet{sokol_etal:15d} (see also Section~\ref{sec:sw}).

In the present study we limit the electron impact ionization to the slow solar wind regime for all latitudes, because we lack appropriate data to differentiate between the slow and fast solar wind electrons. This needs to be a subject of future studies. Limitation to the slow solar wind regime results in overestimation of the electron impact ionization rates for polar regions (see Figures~9 and 10 in \citet{bzowski:08a}). However, due to small fractional input of electron impact ionization to the total ionization rates for the discussed species (see, e.g., Table~\ref{tab:ion}) this simplification does not seriously impact the results discussed, because it is comparable to the accuracy of the total ionization rate model (see the discussion in \citet{bzowski_etal:13b}).

\subsection{Solar wind latitudinal structure \label{sec:sw}}
\begin{figure*}
\begin{tabular}{c}
\includegraphics[width=0.7\textwidth]{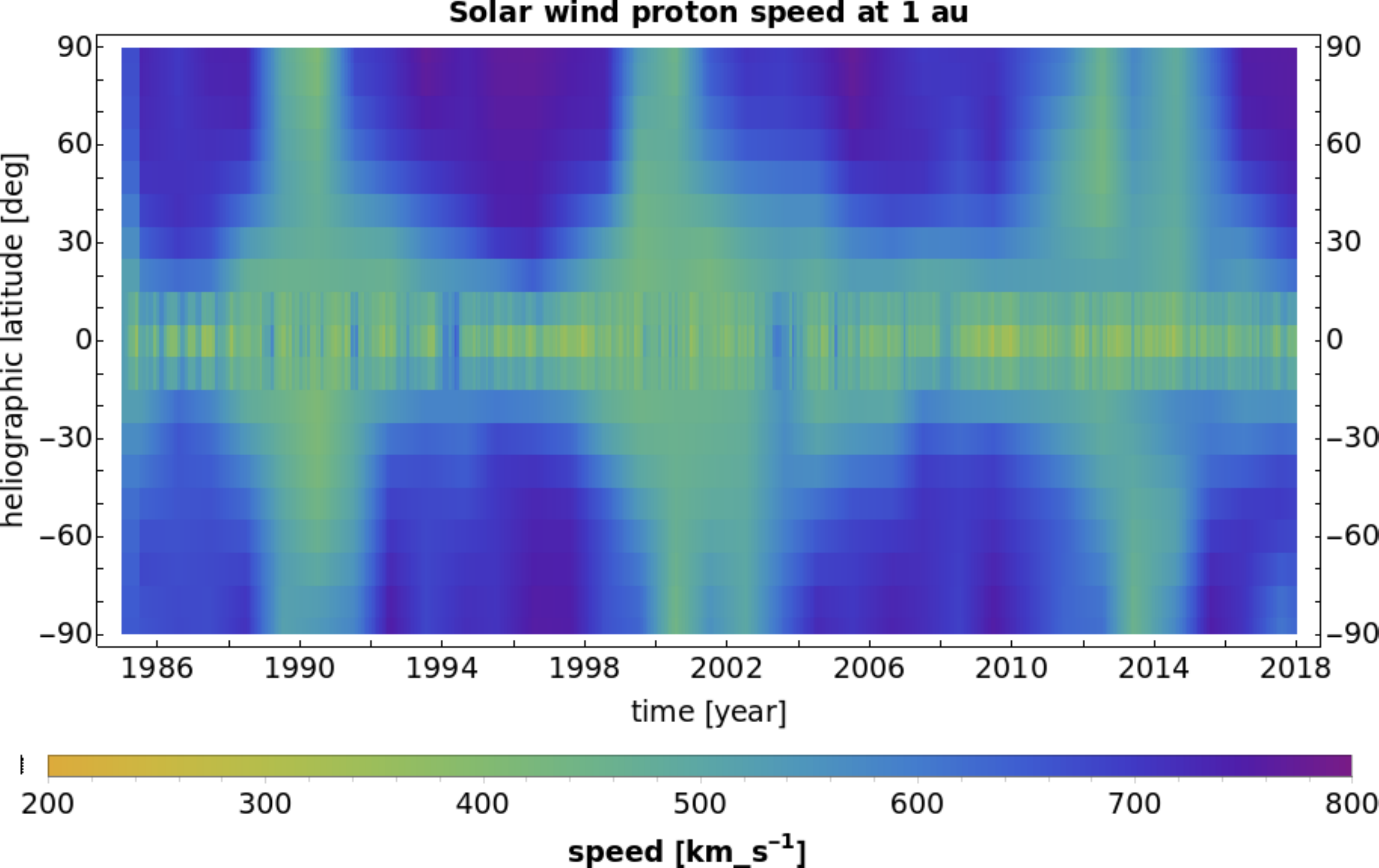}  \\
\includegraphics[width=0.7\textwidth]{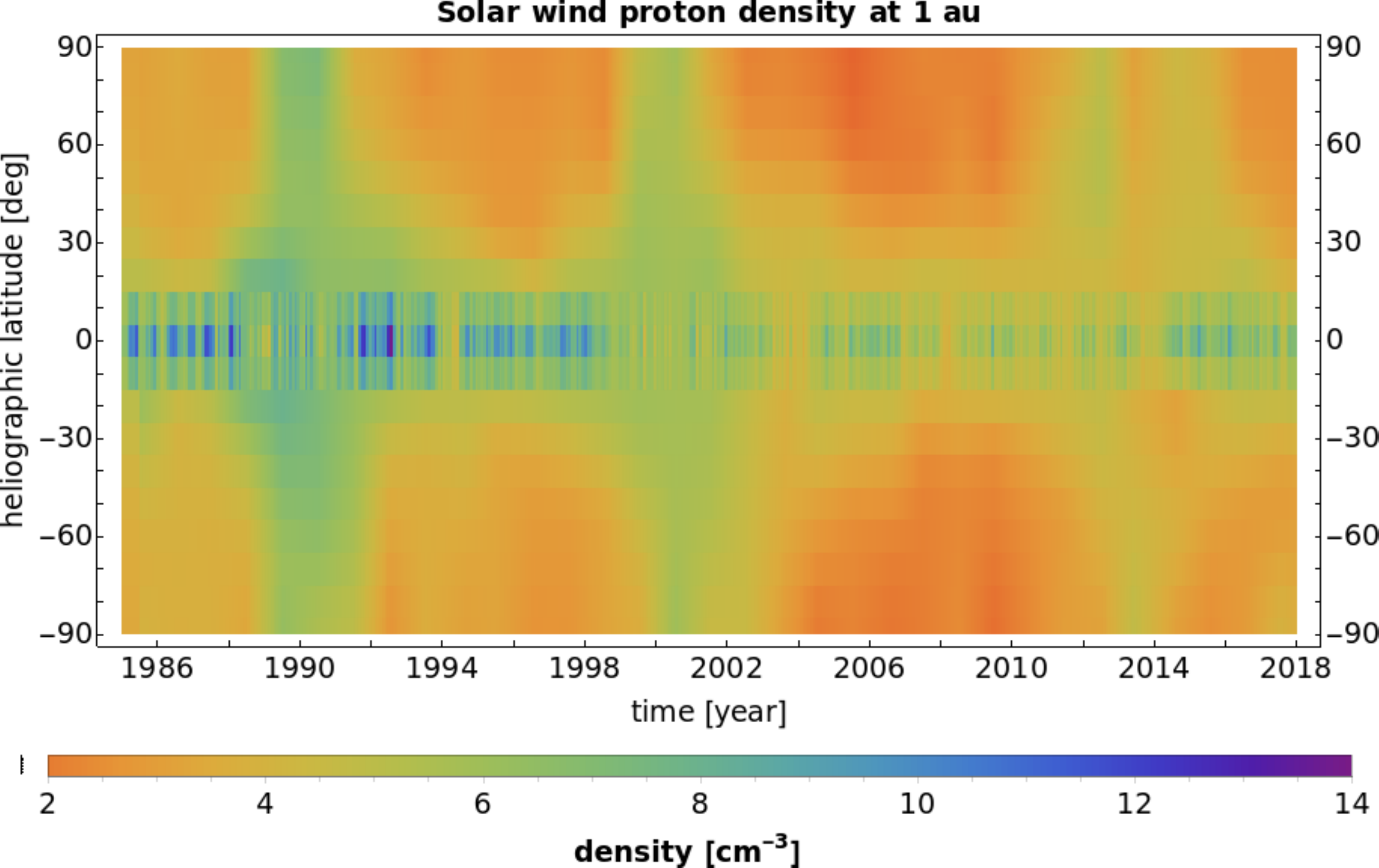}  \\
\end{tabular}
\centering
\caption{Top: solar wind proton speed at 1~au as a function of time and heliographic latitude obtained following the methodology described by \citet{sokol_etal:13a}. Bottom: solar wind proton density at 1~au as a function of time and heliographic latitude calculated based on the methodology described by \citet{sokol_etal:13a} with modification for the method to estimate solar wind density latitudinal variations described in \citet{sokol_etal:15d} and \citet[][Appendix~B]{mccomas_etal:14a} \label{fig:swMap}}
\end{figure*}

Charge exchange and electron impact ionization rates depend on the solar wind speed and density and, in consequence, they vary with heliographic latitude during the solar cycle as the solar wind does (more in Section~\ref{sec:totIonLat}). Additionally, they follow the long-term changes in the solar wind observed in the recent decades (Section~\ref{sec:totIonEcliptic}; \citet{mccomas_etal:08a, sokol_etal:13a, mccomas_etal:18a, tokumaru_etal:18a}). 

We adopt the solar wind structure in and out of the ecliptic plane and its evolution during solar cycle after \citet{sokol_etal:13a}. The methodology described in \citet{sokol_etal:13a} was applied to data covering the time range from 1985 to 2018, as presented in Figure~\ref{fig:swMap}. The time resolution is one Carrington rotation. In the ecliptic plane, the model is based on multispacecraft observations of the solar wind available through the OMNI database \citep{king_papitashvili:05}. For the solar wind structure out of the ecliptic plane, we adopted the solar wind speed retrieved from the IPS observations conducted by the Institute for Space-Earth Environmental Research (ISEE, Nagoya University, Japan; \citet{tokumaru_etal:10a, tokumaru_etal:12b, tokumaru_etal:15a}). The solar wind density is calculated with the use of the speed--density relation as derived from fast orbital scans of \textit{Ulysses}/SWOOPS and from the latitudinal invariant of the solar wind advection energy flux \citep{leChat_etal:12a}, as described in \citet{sokol_etal:15d} and Appendix~B in \citet{mccomas_etal:14a}.

The solar wind data we used cover the last three solar cycles (SC), that is SC~22 (1986 Septemeber --  1996 August), SC~23 (1996 August -- 2008 December), and SC~24 (2008 December -- 2018 April), as presented in Figure~\ref{fig:swMap}. The solar wind proton speed and density is almost uniform in latitude during a short time period during the maximum of solar activity (around 1990, 2001, and 2013-2014), with the slow flow of speed about 450~\kms\, and solar wind proton density being about 9~$\mathrm{cm}^{-3}$ before the decrease in solar wind density in 2000 and about 6~$\mathrm{cm}^{-3}$ after 2000 \citep[see more in, e.g.,][]{mccomas_etal:08a, sokol_etal:13a}. During other phases of solar activity (decreasing, minimum, and increasing), the slow and dense wind flows are restricted to lower latitudes, with limitation to a band of $\pm30\degr$ around solar equator during solar minimum. At the same time, at higher latitudes the flow is fast, with a speed of about 750~\kms\, and with a density about twice lower than in the equatorial band.

\section{Total ionization rates\label{sec:totIon}}
The effective ionization rate for the heliospheric particles is a sum of the rates of all three ionization processes discussed
\begin{equation}
\beta_{\mathrm{tot}}=\beta_{\mathrm{cx}}+\beta_{\mathrm{ph}}+\beta_{\mathrm{el}}.
\label{eq:betaTot}
\end{equation}
All three ionization reactions discussed vary in time, distance to the Sun, and latitude. However, these variations are of different significance for a given ionization process. The charge exchange rate varies significantly as a function of latitude due to the fast/slow solar wind flow variations during the solar cycle (Figure~\ref{fig:swMap}). Also the long-term variations of the solar wind dynamic pressure, weakly correlated with the phase of the solar cycle, are reflected in the charge exchange rates (dark gray lines in Figure~\ref{fig:ionRatesAll}). Photoionization is the ionization reaction that shows the strongest variation with the solar activity, and for which the latitudinal variations need further investigation, as discussed in Section~\ref{sec:ph}. In this study we assumed a moderate variation of photoionization with latitude and thus the variations in time are the main driver of its modulation (dark blue lines in Figure~\ref{fig:ionRatesAll}). However, the latitudinal variations of EUV flux intensity with distance to the Sun may cause our estimate of photoionization rate for distances to the Sun inside 1~au to be underestimated \citep{auchere:05a}. The electron impact ionization (light blue lines in Figure~\ref{fig:ionRatesAll}) varies in latitude as the solar wind does; however, the key parameter for its modulation is its variation with heliocentric distance (Figure~\ref{fig:eleR}). This latter factor makes the electron impact ionization reaction the most significant at close distances to the Sun (within 1~au) and almost negligible elsewhere inside the solar wind termination shock, as illustrated in Figure~\ref{fig:eleRRatio}.

\subsection{In-ecliptic variation \label{sec:totIonEcliptic}}
\begin{figure*}
\begin{tabular}{c}
\includegraphics[width=1.01\textwidth]{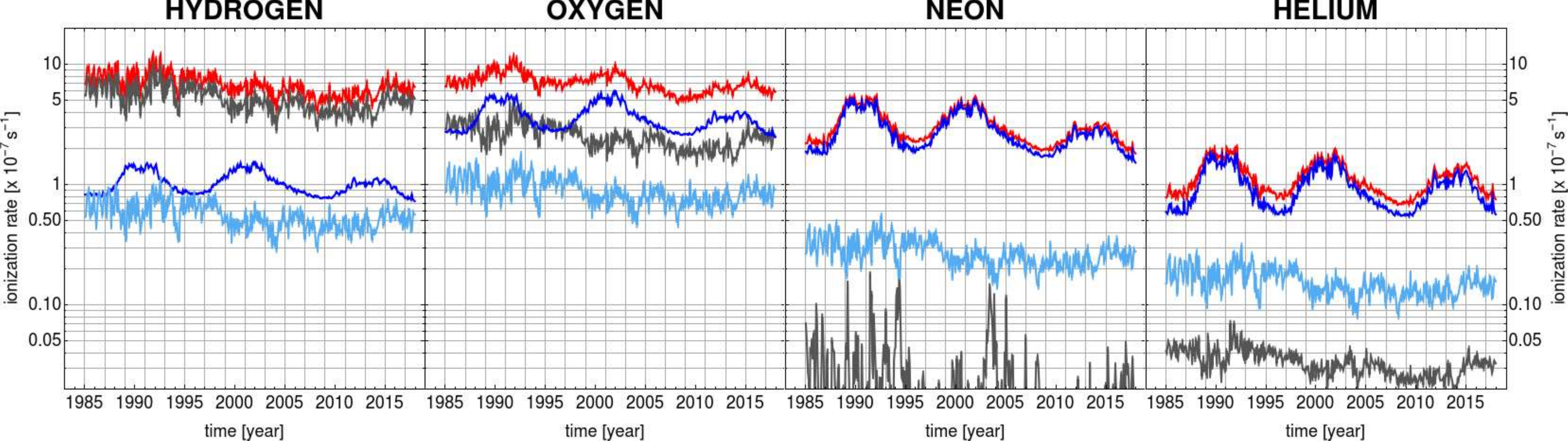}  \\
\includegraphics[width=0.5\textwidth]{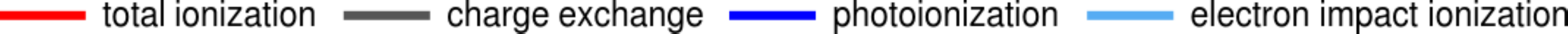}  \\
\includegraphics[width=\textwidth]{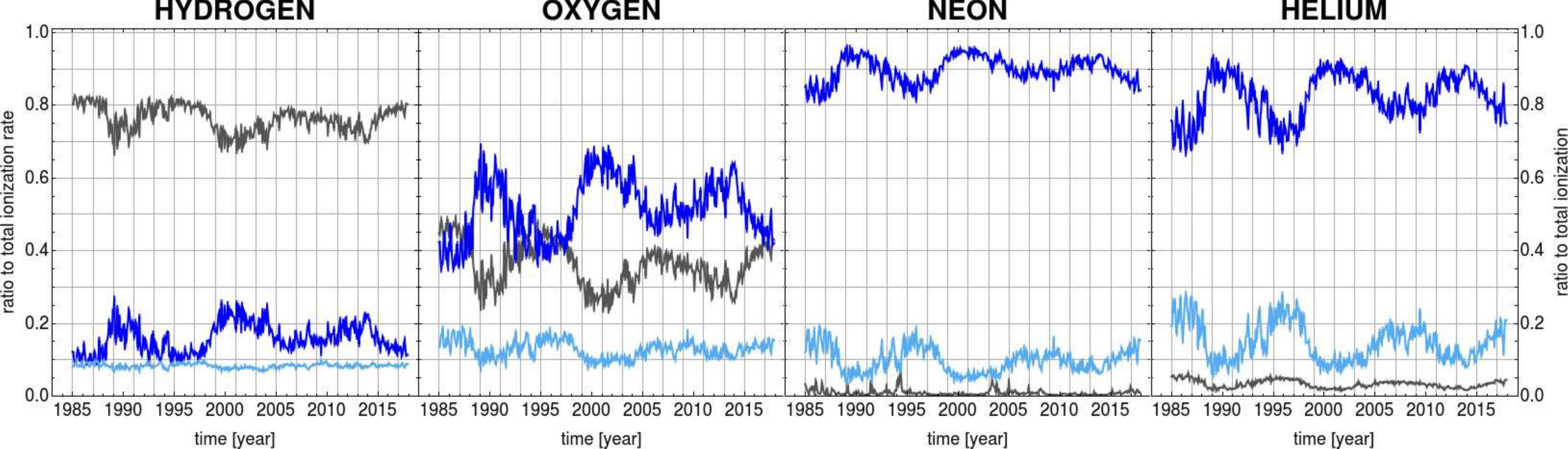} 
\end{tabular}
\caption{Top: time series of ionization rates due to various ionization processes for H, O, Ne, and He in the ecliptic plane at 1~au with Carrington rotation resolution in time. Bottom: time series of the fraction of the individual ionization reaction rates to the total ionization rates for a given species. The color code is given between the two rows of panels. 
\label{fig:ionRatesAll}}
\end{figure*}

\begin{table*}[t]
\centering
\caption{Fraction of Charge Exchange Ionization, Photoionization, and Electron Impact Ionization Rates to the Total Ionization Rates for H, O, Ne, and He at 1~au in the Ecliptic Plane, Averaged Over a Period from 1985 to 2018,  \label{tab:ion}}
\begin{tabular}{r|c|c|c}
  \hline \hline
 \multicolumn{4}{c}{ $\beta_{*}/\beta_{\mathrm{tot}}$: min $\leq$ \textbf{mean $\pm \sigma$} $\leq$ max} \\
  \hline
 \ldots & Charge Exchange & Photoionization & Electron Impact \\
 \hline
 hydrogen & $0.67 \leq $ \textbf{0.76 $\pm$ 0.04} $\leq 0.83$ & $0.09 \leq $ \textbf{0.16 $\pm$ 0.04} $\leq 0.27$ & $0.07 \leq $ \textbf{0.08 $\pm$ 0.01} $\leq 0.10$ \\
 oxygen & $0.23 \leq $ \textbf{0.36 $\pm$ 0.06} $\leq 0.49$ & $0.35 \leq $ \textbf{0.51 $\pm$ 0.08} $\leq 0.68$ & $0.07 \leq $ \textbf{0.13 $\pm$ 0.02} $\leq 0.19$ \\
 neon & $0.001 \leq $ \textbf{0.008 $\pm$ 0.008} $\leq 0.072$ & $0.805 \leq $ \textbf{0.896 $\pm$ 0.035} $\leq 0.958$ & $0.035 \leq $ \textbf{0.096 $\pm$ 0.034} $\leq 0.190$ \\
 helium & $0.01 \leq $ \textbf{0.03 $\pm$ 0.01} $\leq 0.06$ & $0.67 \leq $ \textbf{0.82 $\pm$ 0.06} $\leq 0.93$ & $0.05 \leq $ \textbf{0.15 $\pm$ 0.05} $\leq 0.28$ \\
  \hline
\end{tabular}
\tablecomments{$\sigma$ stands for standard deviation of fractions in the given period. The values in bold are the mean of each range.}
\end{table*}

\begin{table*}[t]
\centering
\caption{Ratios of the total ionization rates for O, Ne, and He with respect to the total ionization rates for H at 1~au in the ecliptic plane and in the polar region, averaged over a period from 1985 to 2018.  \label{tab:ion2}}
\begin{tabular}{r|c|c|c}
 \hline \hline
 \multicolumn{4}{c}{ $\beta_{\mathrm{tot}}\left(*\right) / \beta_{\mathrm{tot}}\left(H\right)$: min $\leq$ \textbf{mean $\pm \sigma$} $\leq$ max} \\
 \hline
 \ldots & oxygen & neon & helium \\
 \hline
 in-ecliptic & $0.82 \leq $\textbf{1.06 $\pm$ 0.14}$ \leq 1.45$ & $0.24 \leq $\textbf{0.47 $\pm$ 0.15}$ \leq 0.94$ & $0.09 \leq $\textbf{0.17 $\pm$ 0.05}$ \leq 0.34$ \\
 polar & $1.09 \leq $\textbf{1.29 $\pm$ 0.11}$ \leq 1.57$ & $ 0.43 \leq $\textbf{0.65 $\pm$ 0.11}$ \leq 0.96$ & $0.14 \leq $\textbf{0.22 $\pm$ 0.04}$ \leq 0.33$ \\
  \hline
\end{tabular}
\tablecomments{The polar ratio is an average for the South and North Poles. $\sigma$ stands for standard deviation of fraction in the given period. The values in bold are the mean of each range.}
\end{table*}

Various ionization processes affect interstellar species in different ways, resulting in different modulations of the heliospheric particles inside the heliosphere and, in consequence, in different ISN gas density distributions and PUI production rates \citep[see, e.g.,][]{sokol_etal:16a}. Most of the available measurements of the heliospheric particles of ISN gas, PUIs, or ENAs are collected by instruments operating in the ecliptic plane and at 1~au. Thus, the variation of the ionization rates close to the ecliptic at 1~au are of high significance for the heliospheric studies and very often serve as a reference for the ionization rates inside the heliosphere. However, because the trajectories of particles measured in the ecliptic plane traverse the heliosphere at different latitudes, the latitudinal variation of ionization rates cannot be neglected, especially in studies of the ISN gas and PUI cone and the downwind hemisphere (see Section~\ref{sec:totIonLat} and discussion in \citet{sokol_etal:16a}).

Figure~\ref{fig:ionRatesAll} compares the in-ecliptic time series of the ionization rates for H, O, Ne, and He at 1~au from 1985 to 2018. All three ionization reactions are presented together with the total ionization rate. The bottom panels of Figure~\ref{fig:ionRatesAll} present the fraction of a given ionization reaction to the total ionization rate. The average values of these fractional contributions over all Carrington rotations in the period from 1985 to 2018 are summarized in Table~\ref{tab:ion}. In addition, we present standard deviations as well as the minimal and the maximal fractional input of an individual ionization reaction to the total ionization rate.

Hydrogen is the most prone to charge exchange with the solar wind protons, which constitutes on average $76\%$ of the total ionization rate for this species (Figure~\ref{fig:ionRatesAll} and Table~\ref{tab:ion}). The remaining ionization reactions for H, photoionization and electron impact ionization, bring up to $16\%$ and $8\%$ to the total ionization losses, respectively. Moreover, ISN~H is significantly affected by the Lyman-$\alpha$ radiation pressure \citep[see][]{kowalska-leszczynska_etal:18a, kowalska-leszczynska_etal:18b}, which has important consequences for the ISN~H and H$^{+}$~PUI distribution.

The other species for which the charge exchange with the solar wind particles is significant is oxygen. Typically, however, charge exchange constitutes only about $36\%$ of its total ionization rate, whereas photoionization brings on average about $51\%$. But, as presented in Figure~\ref{fig:ionRatesAll}, the relative input from the two reactions is strongly variable and can be comparable like, e.g., in 1996 or 2017, which corresponds to the minimum of the solar activity. An exception is the minimum between SC~23 and SC~24, when the solar wind flux was low and the photoionization was responsible for $\sim50\%$ of the ionization for ISN~O, with $\sim 40\%$ due to charge exchange ionization and $\sim10\%$ due to electron impact ionization. The photoionization is the dominant ionization reaction for oxygen during the solar maximum regardless of the solar cycle epoch and is about two times higher than the charge exchange rate. The interplay between these two most effective ionization reactions for ISN~O is modulated in time both by the periodic variations of the solar activity and by the long-term changes in the solar wind dynamic pressure. Additionally, the electron impact ionization rate constitutes on average $13\%$ of the total ionization rates for ISN~O (Table~\ref{tab:ion}), which is almost as much as in a case of ISN~He, and almost twice more than in a case of ISN~H, which is a result of greater cross-sections for oxygen. Furthermore, the absolute electron impact ionization rates are the greatest for oxygen among the four species discussed.

\begin{figure}
\includegraphics[width=0.4\textwidth]{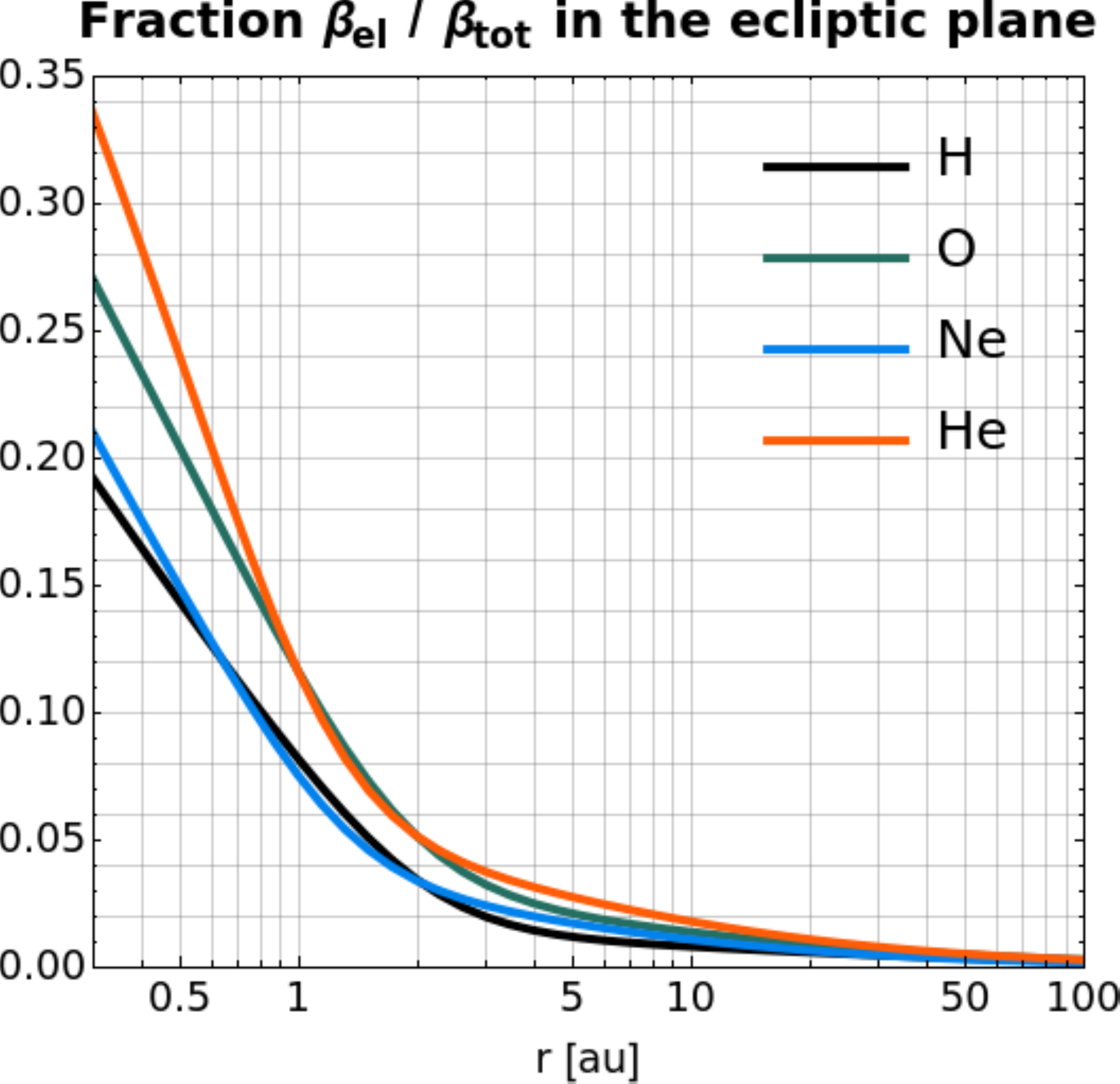}
\centering
\caption{Ratio of the electron impact ionization to total ionization rates in the ecliptic plane as a function of distance to the Sun for four species discussed for a moderate solar activity, as in 1999. \label{fig:eleRRatio}}
\end{figure}

For ISN~He and Ne, the dominant ionization reaction is ionization by the solar EUV radiation (see Figure~\ref{fig:ionRatesAll} and Table~\ref{tab:ion}). Also of importance is electron impact ionization, which is higher for Ne than for He, but brings more to the total ionization rates for He than for Ne. The electron impact ionization for He can contribute to the total ionization rate at 1~au more than $\sim25\%$ during solar minimum, and as low as $\sim5\%$ during solar maximum. This fractional input of electron impact ionization to the total ionization rate for He increases for distances to the Sun smaller than 1~au (Figure~\ref{fig:eleRRatio}). The charge exchange with solar wind particles is almost negligible for these two species. Because the solar EUV flux varies in time with the solar activity, the ionization losses for He and Ne also vary during solar cycle and are two-fold higher during solar maximum than during solar minimum (Figure~\ref{fig:ionRatesAll}).

A comparison among the species shows that in the ecliptic plane the highest ionization rates are for hydrogen and oxygen, for which the total ionization rates are almost identical, as presented in Figures~\ref{fig:ionRatesAll} and \ref{fig:ionRatesTot}, and Table~\ref{tab:ion2}. The lowest ionization rates are for helium, being less than $20\%$ of the total ionization rate for hydrogen. The total ionization rate for neon is about 0.5 of the hydrogen total ionization rate. 

Two groups among the ionization rates can be identified. One that follows the solar cycle variations in time, like those for He and Ne, and the other for which these variations are not directly reflected in the in-ecliptic rates, but are present out of the ecliptic plane, like in the case of H and O. This is because the solar wind in the ecliptic plane does not show a clear quasi-periodic variability related to the cycle of solar activity (Figure~\ref{fig:swMap}). Consequently, the total ionization rate of the species for which the charge exchange ionization dominates, does not show a clear quasi-periodic solar cycle variations in the ecliptic plane (Figure~\ref{fig:ionMaps}). Since solar wind features a clear solar cycle-related modulation in the polar regions, also the charge exchange rate for H and O in the polar regions is modulated in phase with the solar cycle, as further discussed in Section~\ref{sec:totIonLat}.

\subsection{Latitudinal variation \label{sec:totIonLat}}
\begin{figure*}
\begin{tabular}{c}
\includegraphics[width=0.7\textwidth]{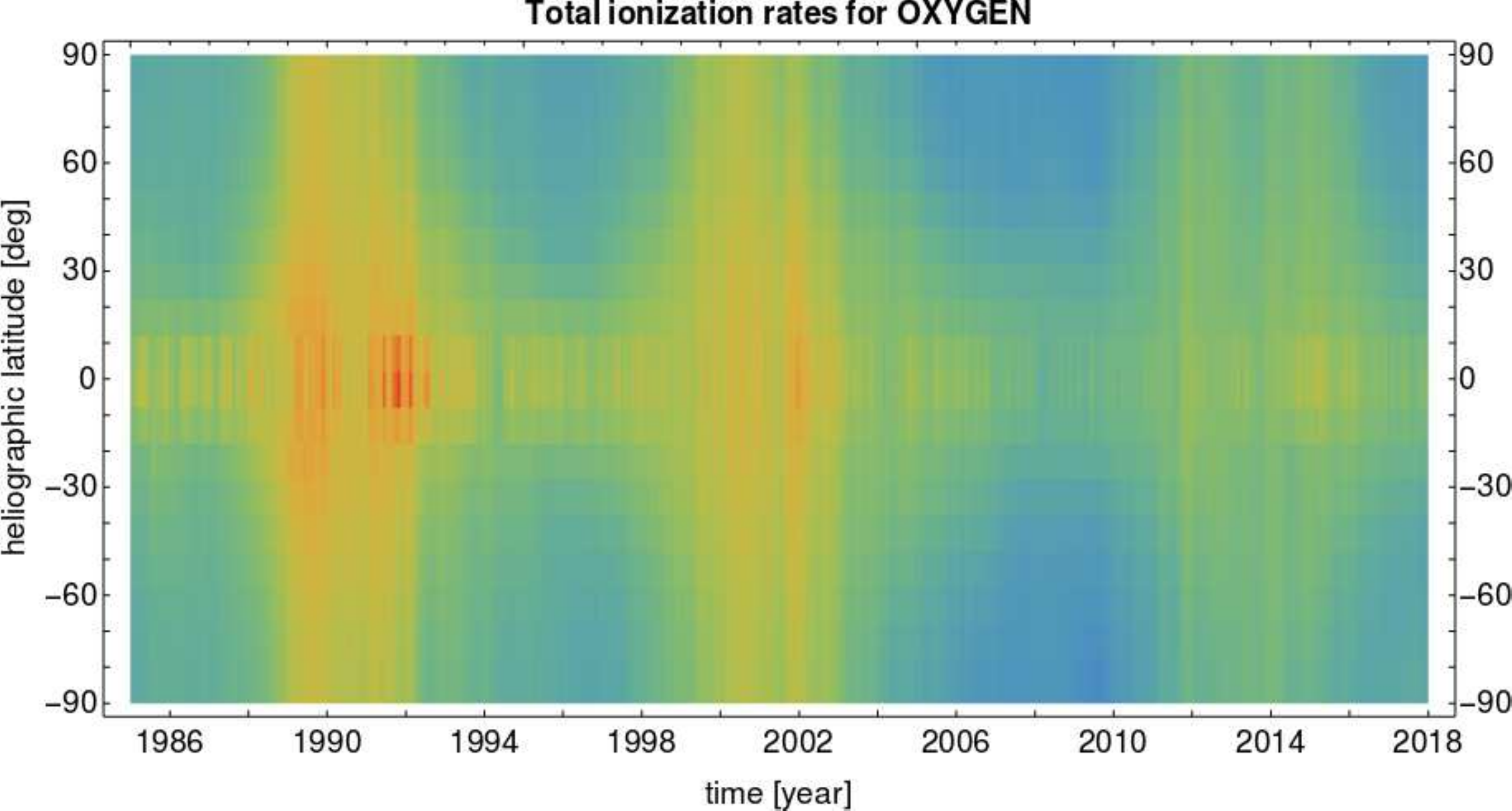} \\
\includegraphics[width=0.7\textwidth]{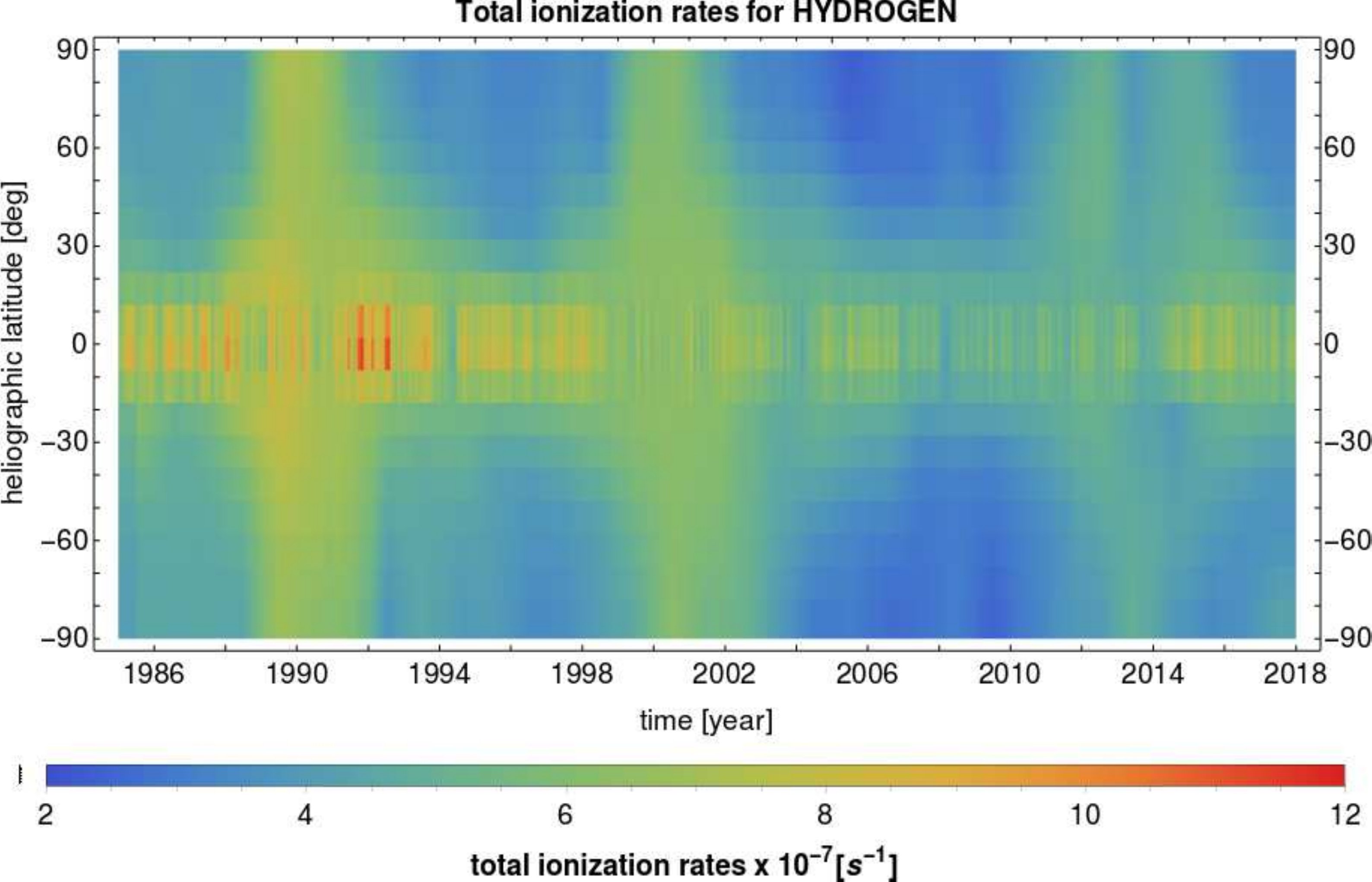} \\
\end{tabular}
\centering
\caption{Total ionization rates at 1~au as a function of heliographic latitude and time for oxygen (top panel) and hydrogen (bottom panel). Color bar is the same for both panels.\label{fig:ionMaps}}
\end{figure*}

\begin{figure*}
\includegraphics[width=\textwidth]{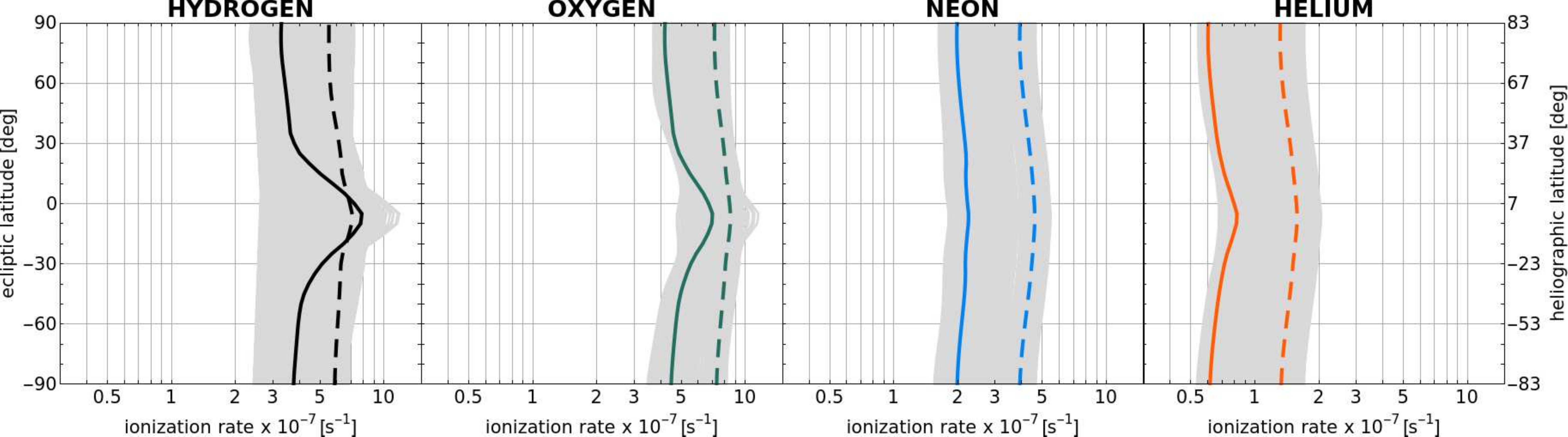}
\centering
\caption{Total ionization rates at 1~au as a function of ecliptic (heliographic) latitude at 1~au during solar minimum ($\sim 1996$, solid line) and solar maximum ($\sim 2001$, dashed line). The gray shadows mark the region occupied by the latitudinal profiles for all Carrington rotations in the period from 1985 to 2018. \label{fig:ionRatesTotLat}}
\end{figure*}

\begin{figure}
\includegraphics[scale=0.35]{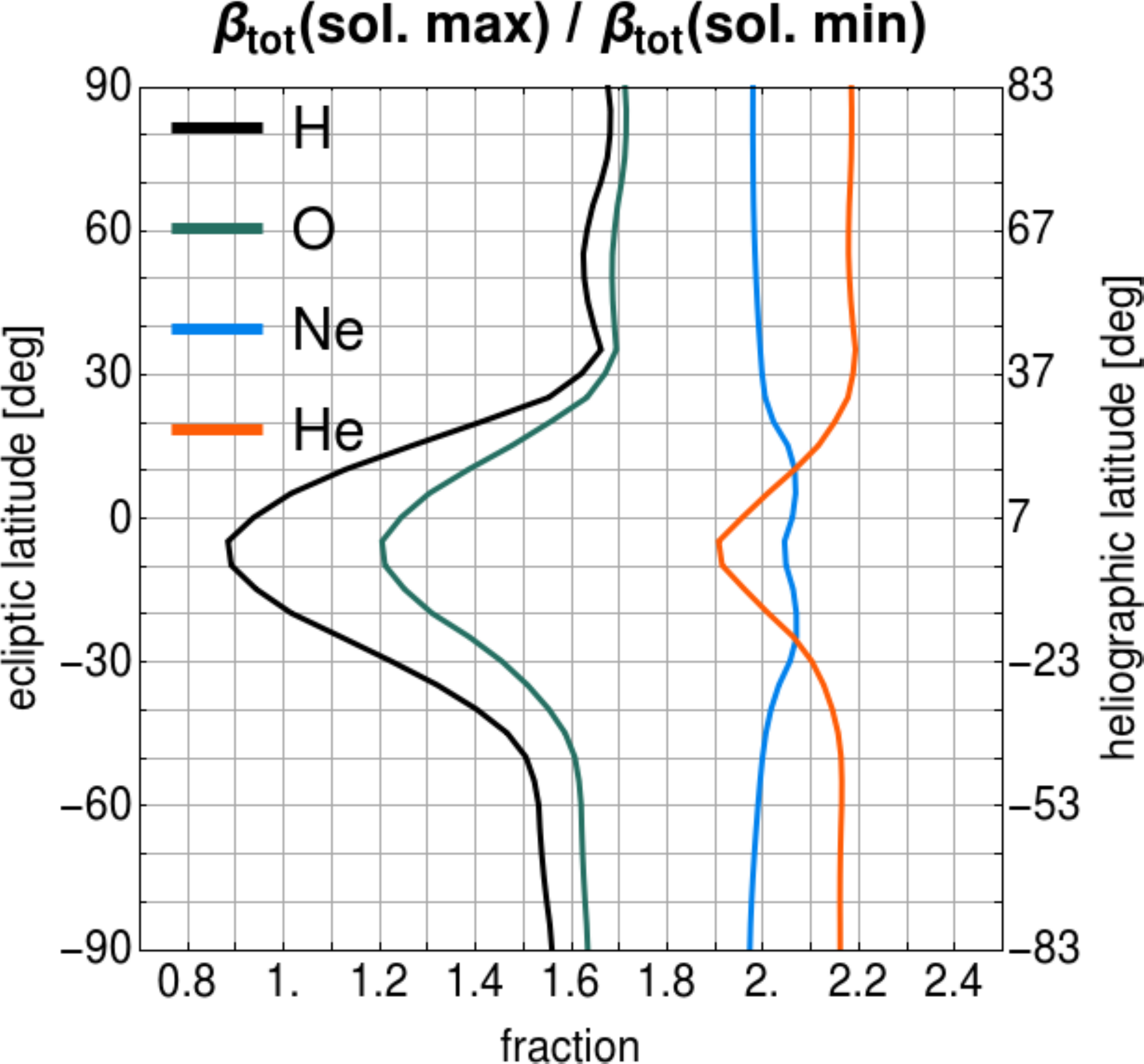}
\centering
\caption{Fraction of the latitudinal profiles of the total ionization rates during solar maximum ($\sim 2001$) to solar minimum ($\sim1996$) for H, O, Ne, and He. \label{fig:ionRatesTotLatRatio}}
\end{figure}

Due to the dependence on the solar activity conditions, the ionization rates vary in latitude during the solar cycle as the solar wind flow and the solar EUV flux does. The charge exchange is almost uniform as a function of latitude during solar maximum as consequence of almost uniform solar wind structure at that time (Figures~\ref{fig:ionMaps} and \ref{fig:ionRatesTotLat}). During solar minimum and phases of decreasing and increasing solar activity, the solar wind has bi-modal structure in latitude, which is reflected in charge exchange and electron impact ionization rates, and, in consequence, also in the total ionization rates for species prone to these two ionization process (Figures~\ref{fig:ionMaps} and \ref{fig:ionRatesTotLat}). Also the distribution of the EUV flux sources varies with latitude during the solar cycle \citep[see, e.g.,][]{auchere_etal:05a}, which may influence the anisotropy of the photoionization rates in time.

For neon and helium, the total ionization rates vary moderately with latitude with  solar activity (Figure~\ref{fig:ionRatesTotLat}). This is because the dominant ionization reaction for these species is photoionization (Table~\ref{tab:ion}), for which we used models with assumed mild latitudinal variations (see Section~\ref{sec:ph}, Equation~\ref{eq:phLat}), which according to the discussion in \citet{auchere:05a,auchere_etal:05a,auchere_etal:05c, bzowski_etal:13a}, might be underestimated and require futher study in the future.

Additionally, for helium, the electron impact ionization may contribute significantly and add latitudinal variations in time to the total ionization rates (Figure~\ref{fig:ionRatesTotLat} and \ref{fig:ionRatesTotLatRatio}). Also, the significance of the electron impact ionization rate strongly increases with the decrease of the distance from the Sun (Figure~\ref{fig:eleR} and \ref{fig:eleRRatio}; Section~\ref{sec:ele}, \citet{bzowski_etal:13a, bzowski_etal:13b, mcmullin_etal:04a}). Therefore, the latitudinal anisotropy of electron impact ionization for all species also increases with the decrease of heliocentric distance during low and moderate solar activity.

Consequently, the total ionization rates vary significantly both in time and in latitude as presented in Figures~\ref{fig:ionRatesTotLat} and \ref{fig:ionRatesTotLatRatio} for selected epochs, and in Figure~\ref{fig:ionMaps} for H and O for a period 1985-2018. According to the adopted models, the total ionization rates for He and Ne are approximately two times higher during solar maximum than during solar minimum (Figure~\ref{fig:ionRatesTotLatRatio}). For H and O, the total ionization rates at mid and high latitudes are about 1.5 higher during solar maximum than during solar minimum (Figure~\ref{fig:ionRatesTotLatRatio}). The in-ecliptic total ionization rates for O vary about $20\%$ during the solar cycle with higher rates during solar maximum. For H, the in-ecliptic variations are about $10\%$ with higher rates during solar minimum, as presented in Figure~{\ref{fig:ionRatesTotLat}} and in Figure~\ref{fig:ionRatesTotLatRatio}. Interestingly, during solar minimum the ionization rates for H close to the solar equator can be greater than during solar maximum (see Figure~\ref{fig:ionRatesTotLat}). This is a result of the long-term changes in the solar wind flux at time scales longer than the solar cycle period, it is the drop in the solar wind proton density after 2000 (see the discussion in, e.g., \citet{mccomas_etal:08a} and \citet{sokol_etal:13a}). The amplitude of pole-to-ecliptic variations with the solar cycle, within the adopted models, is the greatest for hydrogen, and the smallest for neon, as illustrated in Figure~\ref{fig:ionRatesTotLatRatio}. However, this conclusion may change after more thorough study of the latitudinal variations of the photoionization rates become available. Table~\ref{tab:ion2} compares the total ionization rates for various species with respect to total ionization rates for hydrogen. For polar regions the total ionization rates are the greatest for oxygen being almost $30\%$ greater than for hydrogen. 

\section{Summary and conclusions \label{sec:Summary}} 
\begin{figure*}
\includegraphics[width=\textwidth]{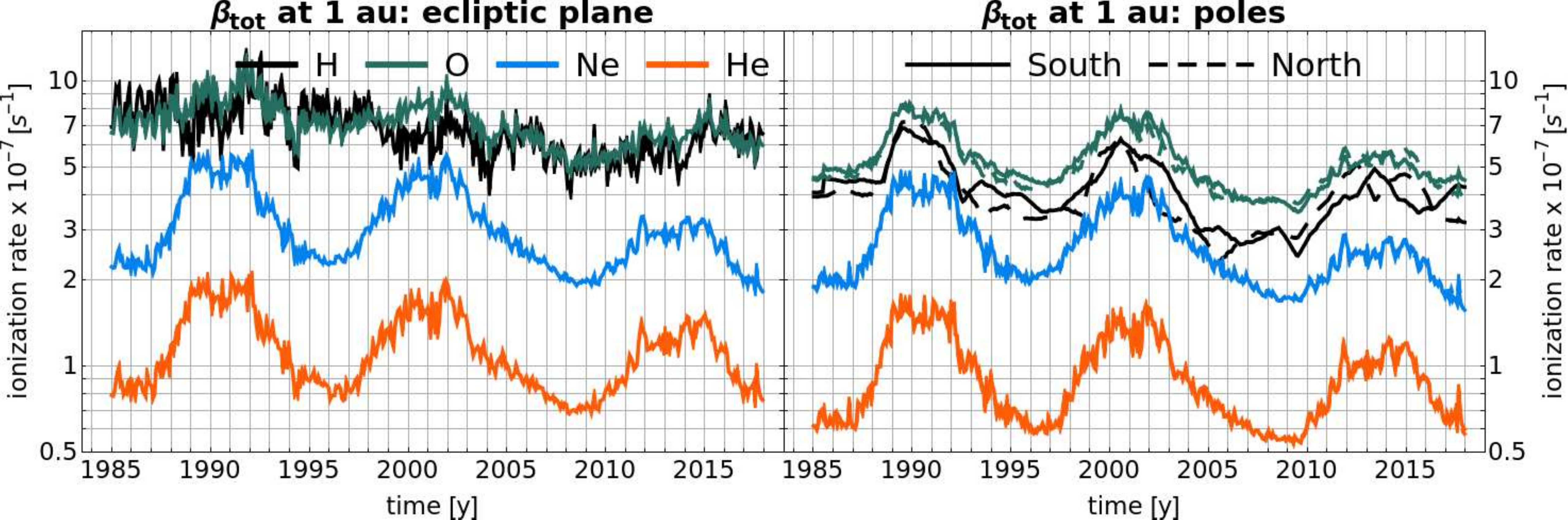}
\centering
\caption{Left: time series of the total ionization rates in the ecliptic plane at 1~au (see the red line in Figure~\ref{fig:ionRatesAll}). Right: total ionization rates at 1~au at ecliptic poles (solid line - South pole, dashed line - North pole; note that the dashed lines are indistinguishable for O, Ne, and He). \label{fig:ionRatesTot}}
\end{figure*}

We discussed the ionization processes relevant for the ISN hydrogen, oxygen, neon, and helium inside the heliosphere estimated based on the published methodology. We focused on modulations in time, with heliocentric distance, and with latitude. We discussed similarities and differences in ionization processes for the given species. We studied the relations between the total ionization rates both in the ecliptic plane and in the polar regions. 

The study shows that for hydrogen and oxygen the solar ionization is the strongest (Figure~\ref{fig:ionRatesTot}), and thus the resulting modulation of the H and O fluxes of heliospheric particles is expected to be the highest. The lowest modulation by solar ionizing factors is for helium, it is almost an order of magnitude smaller than for hydrogen and oxygen at 1~au and in the ecliptic plane. For He and Ne, the main source of ionization losses is photoionization, and thus modulation for these species is correlated with the solar cycle variations. Hydrogen atoms are prone to the solar wind variations both in time and in latitude. Oxygen is a species for which both charge exchange and photoionization losses can be dominant ionization sources depending of the phase of solar activity and long-term changes in the solar wind. The total ionization rates are the highest out of the ecliptic plane for oxygen (Figure~\ref{fig:ionRatesTot}). 

The solar ionizing factors act differently on different heliospheric particles, which results in different modulation of these particles throughout the heliosphere. This brings important consequences for the study of heliospheric particles, like the ISN gas, PUIs, and ENAs, as well as physical processes in the inner and outer heliosphere.

\acknowledgments
The presented study is supported by the Polish National Science Center grant No. 2015/19/B/ST9/01328. The IPS observations were made under the solar wind program of the ISEE.

\bibliographystyle{apj}
\bibliography{sokol_PUIsIonization_accepted_clear_2arxiv.bbl}{}

\end{document}